\documentclass[12pt]{article}

\catcode`\@=11

\def\section{\@startsection{section}{1}{\z@}{3.5ex plus 1ex minus
 .2ex}{2.3ex plus .2ex}{\bf}}

\def\thesubsection{\arabic{section}.\arabic{subsection}}
\renewcommand{\subsection}[1]{\addtocounter{subsection}{1}
\vspace{2.5mm}\par\noindent {\it \thesubsection . #1}\par
 \vspace{0.5mm} }
\catcode`\@=12
\newfont{\mbm}{msbm10 scaled\magstep1}

\def\reflist{\section*{References\markboth
        {REFLIST}{REFLIST}}\list
        {[\arabic{enumi}]\hfill}{\settowidth\labelwidth{[999]}
        \leftmargin\labelwidth
        \advance\leftmargin\labelsep\usecounter{enumi}}}
 \relax

\newcommand{\be}{\begin{equation}}
\newcommand{\ee}{\end{equation}}
\newcommand{\ba}{\begin{eqnarray}}
\newcommand{\ea}{\end{eqnarray}}

\catcode`\@=11

\begin{document}
\begin{titlepage}
\rightline{{LPT-ORSAY 02-38}}
\rightline{{CPHT RR 038.0502}}
\rightline{{Saclay t02/062}}
\rightline{{hep-ph/0205264}}
\vskip 2cm
\centerline{{\large\bf Multiple seesaw at low energy\footnote{Dedicated
to Stefan Pokorski on the occasion of his 60th birthday.}}}
\vskip 1cm
\centerline{Emilian Dudas${}^{\dagger,\star}$ and Carlos 
A. Savoy ${}^{\ddagger}$}
\vskip 0.5cm
\centerline{\it ${}^\dagger$ Laboratoire de Physique Th{\'e}orique
\footnote{Unit{\'e} Mixte de Recherche du CNRS (UMR 8627).}}
\centerline{\it Univ. de Paris-Sud, B{\^a}t. 210, F-91405 Orsay Cedex}
\vskip 0.3cm 
\centerline{\it ${}^\star$ Centre de Physique Th{\'e}orique,
Ecole Polytechnique, F-91128 Palaiseau}
\vskip 0.3cm 
\centerline{\it ${}^\ddagger$ Service de Physique Th{\'e}orique,
CEA/DSM/SPhT, CEA/Saclay \footnote{Unit\'e de recherche associée au CNRS.}}
\centerline{\it F-91191 Gif-sur-Yvette Cedex, France}

\vskip  1.0cm
\begin{abstract}
A new mechanism for generating neutrino masses without a high-energy
mass scale is proposed. The mechanism needs a fundamental mass scale M
in the 100-1000 TeV region and a minimal field content beyond the
Standard Model one containing a pair of fermion singlets and a pair of
weak doublet fermions for each neutrino mass, all of them with a mass
of order M. The neutrino mass appears by a multiple seesaw-type
tree-level diagram. We provide an explicit model based on supersymmetry
and an abelian symmetry which provides the required fermion mass
matrix. The mechanism is natural in the context of string theories with
a low fundamental scale.  Within an explicit example where the abelian
symmetry is also responsible for the generation of fermion masses and
mixings, we give a hint relating the fermion mass matrices and the weak
mixing angle. By assuming the weak-strong couplings unification, one 
naturally finds $\sin^2 \theta_w = 1/4$ at the fundamental scale.
\end{abstract}
\end{titlepage}

%%%%%%%%%%%%%%%%%%%%%%%%%%%%%%%%%%%%%%%%%%%%%%%%%%%%%%%%%%%%%%%%%%
\section{The minimal model}

Let us consider one active neutrino in a theory whose fundamental mass
scale $M$ is low, of the order of $100-1000 \ TeV$ \cite{ads,ddg}. In
such a context, the smallness of the neutrino masses cannot be
explained by the standard seesaw mechanism \cite{seesaw}, which would
give much too large values. A successful generation of neutrino masses
asks in this case for two conditions. The first is to forbid the
operator responsible for the seesaw mass $(1/M) \ H H \nu_L \nu_L$. The
second is to generate a small neutrino mass by some other mechanism.

Our proposal involves a minimal set of two Standard Model singlet fermions $\Psi_1,
\Psi_2$ and two $SU(2)_L$ doublets $\Psi_4, \Psi_3$ of hypercharges $Y
= \pm 1/2$, respectively.  The symmetries of the model, whose
discussion (in a particular realisation) is postponed to the next
section, give mass terms in the lagrangian of the form
\be
{1 \over 2} \ V \ {\cal M} \ V^T \ , \label{m1}
\ee
where the vector $V$ denotes the collection of fermionic fields
$V = (\nu_L, \Psi_1, \Psi_2, \Psi_3, \Psi_4)$.
The mass matrix ${\cal M}$ in (\ref{m1}) is given by  
\be
      {\cal M} ~=~ \pmatrix{
         0 &  m_1   &   0 &   0  &   0  &  \cr
         m_1 &  M_1 &   M_2  &   m_2  &   m_3  & \cr
         0 &  M_2   &  0  &   m_4  &   m_5  &  \cr
         0 &  m_2   & m_4  & 0  & M_3  & \cr
         0 &  m_3   & m_5  & M_3  &  0  &  \cr }~ \ .
\label{m2}
\ee

The entries $m_i$ denote electroweak-type mass terms, given by the
vacuum expectation value(s) of the Higgs field(s), whereas the capital
letters $M_a$ denote Majorana mass terms (independent of the electroweak  
scale), generically of the order of the fundamental scale, in our case
assumed to be of the order $100-1000 \ TeV$. The eigenvalues/eigenvectors
of the mass matrix (\ref{m2}) are obtained rather easily in the
relevant approximation $M_a >> m_i$. In this limit, the mass matrix has
a block-diagonal form of a $3 \times 3 $ matrix times a $2 \times 2$  
mass matrix. There are  four heavy eigenstates of mass approximately
given by
\ba
&& \lambda_{2,3} \simeq {1 \over 2} \ (M_1 \pm \sqrt{M_1^2 + 4 M_2^2}) \ , 
\nonumber \\
&& \lambda_{4,5} \simeq \ \pm M_3 \ . \label{m3}
\ea 
In this block-diagonal limit, the lightest neutrino stays massless, despite
the presence of the Majorana mass term $M_1 \Psi_1 \Psi_1$ in
(\ref{m2}). This is easily explained by considering the $3 \times 3$
upper block in (\ref{m2}), which has zero determinant. Indeed,
by inverting the $2 \times 2$ matrix block for the singlets 
$\Psi_1, \Psi_2$ 
\be
      {\cal M}_2 ~=~ \pmatrix{
         M_1 &  M_2  &  \cr
         M_2 &  0  &  \cr }~ \ ,
\label{m4}
\ee
we get in the seesaw diagram entries in the singlet propagators of the
type $ S_{12} (p=0) = (1/M_2) $ and $ S_{22}(p=0) = (M_1/M_2)^2$, where
$p$ is the momentum flowing into the fermion propagator. The main point
here is that the $S_{11}(p=0)$ entry {\it is zero}. Consequently, the
usual seesaw diagram obtained by integrating out the heavy $\Psi_1 $
field is forbidden. The neutrino mass appears only by taking into
account the small $3 \times 2$ block off-diagonal terms in (\ref{m2}).
In this case, the light eigenvalue is most easily obtained by computing
the determinant of the matrix (\ref{m2})
\be
\det {\cal M} \ = \ - 2 \ m_4 \ m_5 \ m_1^2 \ M_3 \ . \label{m5}
\ee
By combining (\ref{m3}) and (\ref{m5}), we obtain the value of the
light neutrino eigenstate, which in our simplified model is mostly 
given by the electron neutrino 
\be
m_{\nu} = \lambda_1 \simeq \ - {2 m_1^2 m_4 m_5 \over M_2^2 M_3} \ . \label{m6}
\ee
As expected, the corresponding eigenvector $|\hat{\nu}_L >$ is mainly composed
of the lightest neutrino state. The explicit expression is given by
\be
|\hat{\nu}_L> \, \simeq \ |\nu_L> - \, {2 m_1 m_4 m_5 \over M_2^2 M_3}|\Psi_1> 
-\, {m_1 \over M_2} |\Psi_2> + \, {m_1 m_5 \over M_2 M_3} |\Psi_3>
+ \, {m_1 m_4 \over M_2 M_3} |\Psi_4> \ . \label{m06}  
\ee

The final result (\ref{m6}) can be easily understood as a result
of a multiple seesaw-type diagram obtained by integrating out the heavy
fields $\Psi_1, \Psi_2, \Psi_3, \Psi_4$ and inserting the appropriate
Higgs vev(s). The first step is to integrate out the
weak fermion doublets $\psi_3,\psi_4$. The Majorana mass matrix
(\ref{m4}) is then modified to
\be
      {\cal M}_2 ~=~ \pmatrix{
         M_1 &  M_2  &  \cr
         M_2 &  {2 m_4 m_5 \over M_3}  &  \cr }~ \ .
\label{m7}
\ee
The second step involves integrating out the singlet $\psi_1,\psi_2$
fermion fields in the remaining $3 \times 3$ mass matrix
\be
      {\cal M}_3 ~=~ \pmatrix{
      0 &  m_1  & 0 & \cr
      m_1 &  M_1  & M_2 & \cr   
      0 & M_2 & {2 m_4 m_5 \over M_3}  &  \cr }~ \ ,
\label{m8}
\ee
leading to the eigenvalues $\lambda_{2,3}$ in (\ref{m3}) and $\lambda_1
= m_{\nu}$ in (\ref{m6}). 

The lagrangian leading to the mass matrix
(\ref{m2}) breaks the $U(1)$ lepton number which we define to act on the
various fields as
\ba
&& \nu_L \rightarrow e^{i \alpha} \nu_L \ , \psi_1 \rightarrow  
e^{-i \alpha} \psi_1 \ , \ \psi_2 \rightarrow  
e^{i \alpha} \psi_2 \ , \nonumber \\
&& \psi_3 \rightarrow e^{-i \alpha} \psi_3 \ , \ \psi_4 \rightarrow  
e^{-i \alpha} \psi_4 \ . \label{m9}
\ea  
The breaking of the lepton number (by $\Delta L=2$) is due to the 
Majorana mass parameters $M_1$ and $M_3$. However, only $M_3$ is
relevant for the neutrino mass generation. The smalness of the neutrino mass
(\ref{m6}) is then understood by the fact $\psi_3$ and $\psi_4$ do not
directly couple to the lightest neutrino. The transmission of the
lepton number breaking goes therefore through the heavy fermions
$\psi_1,\psi_2$ and generates a cubic supression in the heavy masses 
(\ref{m6}). 

We now add the usual mass for the electron and a new mass mixing
$\psi_3^{+}$ and the right-handed electron in the effective theory.
The charged lepton mass matrix in the theory is of the form
\ba 
\left( e_L^c \Psi_3^{+}
\right) 
\left(
\begin{array}{cc} 
m_6 & 0 \\ 
m_7 & M_3
\end{array}
\right)
\left(
\begin{array}{c}
e_R \\
\Psi_4^{-}
\end{array}
\right) \ ,\label{m09}
\ea
and the physical electron mass is approximately given by $m_e \sim m_6$ ,
while the charged new leptons have a mass of order $M_3$.
The physical charged states mix slightly the light and the heavy states.
The physical electron state is for example given by
\be
|{\hat e}_R> \ \simeq \ |e_R> - {m_7 \over M_3 } |\Psi_4^{-}> \ , \ \ \ \
|{\hat e}_L> \ \simeq \ |e_L> - {m_7 m_6 \over M_3^2 } |\Psi_3^{-}>
\ . \label{m091}
\ee

With the minimal field content (\ref{m2}) only one neutrino linear
combination acquires a mass at tree level. In order to give a tree-leel
mass of the type (\ref{m6}) for a second neutrino, we need a second
pair of singlet fields $\psi_5, \psi_6$ and a second pair of weak
fermion doublets $\psi_7,\psi_8$. The second light neutrino should couple to
$\psi_1$ and $\psi_5$, but not to $\psi_2$ and $\psi_6$.  The need of
doubling the exotic fermion spectrum can be easily shown by following
the integrating out procedure outlined above. Indeed, first of all we
need to double the singlet fermion content in order to provide two
(linear combinations of) light neutrinos to couple to two different
singlets. On the other hand, by integrating out the weak doublets we
generate the Majorana mass matrix for the singlet fields. With only one
pair of weak doublets, the Majorana mass matrix for the singlets will
have zero eigenvalues and as a result one zero mass eigenstate will
survive in the physical spectrum. Adding a second pair of weak doublets
will solve this problem by generating large Majorana masses for all
singlets $\psi_1,\psi_2,\psi_5,\psi_6$. The charge lepton mass matrix
in this case is a straightforward generalization of (\ref{m09}). The
light charged states get mixed with heavy charged leptons, analogously
to (\ref{m091}). By promoting the mass entries in (\ref{m2}) and
(\ref{m09}) to $3\times 3$ matrices in the flavour space, all the three
light neutrinos get masses.

The mixing between the light and the heavy states (\ref{m06}) and (\ref{m091})
generates contributions to the anomalous magnetic moment of the electron and muon,
the electric dipole moment and $\mu \rightarrow e \gamma$. For values
of the fundamental scale $ 100 \ TeV < M < 1000 \ TeV$, the mixings are
however very small and the results of such processes are experimentally
unobservable, independently of the mixing angles between the light
neutrino flavors.

There already exist in the literature other mechanisms \cite{ddg2} for getting
small neutrino masses in the presence of large extra dimensions in
models with a low string scale \cite{ads,ddg}. The mechanism we put
forward in our paper does not rely crucially on large extra
dimensions.  Its phenomenology is different in nature from the previous
ones and future experiments could distinguish it from other
mechanisms.
  
The model just presented was based on a minimal field content and a
specific mass matrix (\ref{m2}), which contained a certain number of 
important zeros. Their presence must be assured by some symmetry
of the underlying theory. We were unable to find such a symmetry for
an appropriate non-supersymmetric extension of the Standard Model. The
simplest example we found has low-energy supersymmetry and an
additional abelian symmetry, to be explained in the next
section.

%%%%%%%%%%%%%%%%%%%%%%%%%%%%%%%%%%%%%%%%%%%%%%%%%%%%%%%%%%%%%%%%%%
\section{A supersymmetric example based on an abelian symmetry.} 
 
The goal of this section is to present an explicit example of a symmetry
which produces a mass matrix of the form (\ref{m2}). It will also 
forbids radiative corrections to generate the usual seesaw neutrino mass
by symmetry arguments. The model we consider is the minimal supersymmetric 
standard model (MSSM), enlarged with two additional superfield singlets
$\phi_1, \phi_2$ and two superfields $\phi_3, \phi_4$ of $SU(2)_L \times
U(1)_Y$ quantum numbers $({2,-1/2})$, $({2,1/2})$. Their fermionic components
are the heavy fermion fields present in the previous section. The model
is therefore the minimal supersymmetrization of the 
previous one. The symmetry which will guarantee the existence of the mass
matrix (\ref{m2}) is an abelian $U(1)_X$ symmetry. For the 
purpose of the present section , its nature (global or local, family 
dependent or independent) is irrelevant. We will come back later on its
interesting implications if the symmetry is local.
There are several possible charge assignements which fulfill our
requirements and we just present here one particularly interesting
example. We denote by small letters
$h_1,h_2,l...$ the $U(1)_X$ charges of the MSSM fields $H_1,H_2,L
\cdots$ and by $x_1 \cdots x_4$ the $U(1)_X$ charges of the heavy superfields
$\phi_1 \cdots  \phi_4$. As usual, we assume the existence of the
(super)field $\Phi$ singlet under the Standard Model of $U(1)_X$ charge
$-1$, whose scalar v.e.v.  spontaneously breaks the abelian symmetry.
We also assume, for reasons related to proton stability and lepton
flavor violation, the existence of R-parity under which the heavy
superfields $\phi_1 \cdots  \phi_4$ have a matter-type parity. 

We consider the charge assignements:
\be
h_1=4 \ , \ h_2 = 0 \ , \ l = -1 \ , \ e = 0 \ , \nonumber \\
x_1=x_3= 2 \ , \ x_2=x_4= - 2 \ . \label{e1} 
\ee
The relevant terms in the superpotential of this model, compatible with
holomorphicity and the charge assignement (\ref{e1}) are
\ba
&& W = \lambda_1 \ ({\Phi \over M}) \ L H_2 \phi_1 + 
\lambda_2 \ ({\Phi \over M})^4 \ H_2 \phi_1 \phi_3  
+  \lambda_3 \ ({\Phi \over M})^4 \ H_1 \phi_1 \phi_4 \nonumber \\
&+& \lambda_4 \ H_2 \phi_2 \phi_3 + \lambda_5 \ H_1 \phi_2 \phi_4 
+ M'_1 \ ({\Phi \over M})^4 \ \phi_1 \phi_1 +  
M'_2 \ \phi_1 \phi_2 \nonumber \\
&+&  M'_3 \ \phi_3 \phi_4 +  \mu' \ ({\Phi \over M})^4 H_1
H_2 +  \lambda_6 \epsilon^3 L E H_1 + \epsilon^6 \lambda_7 \Phi_3 E H_1
\ , \label{e2}
\ea 
where $M$ is the fundamental mass scale of the model. 
We denote the various v.e.v.'s by ${\langle \Phi \rangle / M} \equiv
\epsilon << 1$, $\langle H_{1,2} \rangle \equiv v_{1,2}$. Due to their
large supersymmetric masses in (\ref{e2}), assumed to be much larger than that the
supersymmetry breaking scale, the scalar components of $\phi_1 \cdots
\phi_4$ get no v.e.v's. Then the fermionic
mass matrix derived from (\ref{e2}) is precisely of the form (\ref{m2}),
with the various mass parameters being equal to
\ba
m_1 &=&  \lambda_1 \ \epsilon \ v_2 \ , \ m_2 =  \lambda_2 \ \epsilon^4 v_2 \ ,
\ m_3 =  \lambda_3 \ \epsilon^4 v_1 \ , \ m_4 =  \lambda_2 v_2 \ , \
m_5 =  \lambda_5 v_1 \ , \nonumber \\
M_1 &=&  M'_1 \ \epsilon^4 \ , \ M_2 =  M'_2 \ , \ M_3 =  M'_3 \ , \   
\mu = \mu' \ \epsilon^4 \ . \label{e3}
\ea
The natural values for the ``fundamental'' couplings in (\ref{e2}) are
$\lambda_i \sim 1$ and $M'_i , \mu' \sim M$. In this case, by applying the
expression (\ref{m6}), we obtain the neutrino mass
\be
m_{\nu} \sim \ \epsilon^2 \ { v_2^3 v_1 \over M^3} \ . \label{e4}
\ee
For low values of $\tan \beta = v_2/v_1$ and by taking as a particular
example the value $\epsilon \simeq 0.22$ motivated by considering the
abelian symmetry $U(1)_X$ as responsible for the generation of fermion
masses and mixings (see next section) , we find from (\ref{e4}) for the 
electron neutrino $M \sim 500 \ TeV$. By a slight change of charge 
assignements we can
also consistently find $100 \ TeV < M < 1000 \ TeV$. Notice that the charge
assignement (\ref{e1}) generates an effective $\mu$-term in (\ref{e3})
of the order $(1/400) \ M$, which is therefore in the electroweak energy
range. Consequently, the model automatically produces a successful
$\mu$-term in the low-energy theory. 
 
In this model the absence of the seesaw operator $L L
H_2 H_2$ after integrating out the vector-like heavy states is not an
accident. This operator has $U(1)_X$ charge $-2$ and 
is forbidden by holomorphicity in the superpotential\footnote{Actually,
non-analyticity in $\epsilon$ of the type $1 / \epsilon^n$ can be found 
in the superpotential if chiral-type, with respect of $U(1)_X$, 
heavy states are integrated out. A sufficient condition to obtain
analytic superpotential in $\epsilon$ is to integrate out vector-like
heavy states, which is indeed the case under consideration.}. 
On the other hand,
the neutrino mass (\ref{e4}) obtained by integrating out the heavy
states is described by the higher-dimensional operator
\be
{\epsilon^2 \over M^3} \ L L H_2 H_2 \ ( H_1 H_2 ) \ , \label{e5}
\ee  
of $U(1)_X$ charge $2$ that can appear in the superpotential. As we just
proved, this operator is indeed produced at tree-level by a diagram
containing three propagators of massive states. The charge
of the operator (\ref{e5}) explain also the $\epsilon^2$ factor in (\ref{e4}).  
Notice also that the imposition of R-parity and assignement of matter
parity for the heavy fields $\phi_1 \cdots \phi_4$ forbids in the
superpotential operators of the type
\be
M_5 \ ({\Phi \over M})^2 \ H_2 \phi_3 \ + \ 
M_6 \ ({\Phi \over M})^2 \ H_1 \phi_4 \ . \label{e6}  
\ee
Operators of the type (\ref{e6}), if present, would generate mixings of
the heavy leptons with higgsinos which would ask for a more careful
treatment of the resulting ($7 \times 7$) mass matrix. It is not obvious
that such terms would destroy the prediction for the neutrino
mass (\ref{e4}) or  create other serious problems, but we choose here
the simple option of eliminating them by imposing R-parity. 

The lepton flavor violating processes 
in our model are of two types. The
first are typical of supersymmetric models and constrains as usual 
the sparticle spectrum.
The second type are generated by the new lepton violating sector 
$\Phi_1 \cdots \Phi_4$ in the superpotential (\ref{e2}). 
They put a lower bound on the fundamental scale, which is however 
largely satisfied in our class of models by imposing phenomenologically
relevant values for neutrino masses.
 
%%%%%%%%%%%%%%%%%%%%%%%%%%%%%%%%%%%%%%%%%%%%%%%%%%%%%%%%%%%%%%%%%% 
\section{Electroweak angle and Froggatt-Nielsen mechanism at low energy}

Up to now the details of the abelian flavor symmetry were irrelevant
for producing the multiple seesaw mechanism we are proposing. 
We are tempting now to go one step further and try to use the $U(1)_X$
symmetry as a horizontal symmetry responsible for the generation of
fermion masses and mixings \cite{fn}. The quark and charged lepton mass
matrices in this framework can be related to mixed anomalies of $U(1)_X$ 
with the Standard Model gauge group \cite{ir}. Let us denote by $A_1,A_2,A_3$
the mixed anomalies $[U(1)_Y]^2 U(1)_X$, $[SU(2)_L]^2 U(1)_X$ and   
$[SU(3)_c]^2 U(1)_X$ respectively. A straighforward generalization of
the results of \cite{ir} give the relations between the determinants of 
the up quark $Y_U$, down quark $Y_D$ and charged leptons $Y_L$ mass matrices
\ba
\det (Y_D^{-2} Y_L^2) &=& \epsilon^{A_1+A_2-{8 \over 3} A_3-2(h_1+h_2)-
2 (x_3+x_4)} \ , \nonumber \\
\det (Y_U Y_D^{-2} Y_L^3) &=& \epsilon^{ {3 \over 2} (A_1+A_2-2 A_3)-
3 (x_3+x_4)} \ . \label{w1}
\ea

Let us define the number $x = h_1+q_3+d_3 = d_1 + l_3 + e_3$ related to
the angle $\tan \beta = (v_2/v_1)$ approximately by the relation 
$\tan \beta = (m_t/m_b) \ \epsilon^x$. Phenomenologically relevant values
of $x$ are then $x=0,1,2$. Considering the most viable mass matrices 
obtained in the references \cite{ns,ir} we find the relations
\ba
A_1+A_2-2A_3 &=& 12 + 2x + 2 (x_3+x_4) \ , \nonumber \\
A_1+A_2-{8 \over 3} A_3 &=& 2 (h_1+h_2 + x_3+x_4) \ . \label{w2}
\ea
If we insert the charges (\ref{e1}) and consider the large $\tan \beta$ regime 
$x=0$, we naturally find a solution to the eqs. (\ref{w2})
\be
A_1 \ = \ 3 A_2 \quad , \quad A_2 = A_3 \ . \label{w3}  
\ee
In the regime $\tan \beta \sim 1$ , we also find the solution
(\ref{w3}) provided that we change slightly the $H_1$ higgs $U(1)_X$ charge to
$h_1 = 5$. The relation (\ref{w3}) provides an interesting hint for the
value of the weak mixing angle in such a model with low fundamental
scale \cite{ads}. Indeed, in an effective string theory context, the values of the
mixed gauge anomalies can be related to the value of the weak mixing
angle at the energy scale where the flavor symmetry is spontaneously
broken. The relevant string theory under consideration is here Type I
string theory with low fundamental mass scale $M_I \sim M$.
The $SU(2)_L$ and $U(1)_Y$ gauge groups leave on a stack of
``electroweak'' branes. The tree-level gauge couplings in this case are 
given by (we consider for illustration D9 branes) \cite{iru}
\be
{4 \pi^2 \over g_a^2 } = s + s_{ak} m_k \ , \label{w4}
\ee
where $s = Re S $ is the four-dimensional dilaton and $m_k = Re M_k$ are
twisted moduli present in orbifold-type compactifications. Under the
$U(1)_X$ gauge transformation the twisted moduli get shifted according
to
\ba
&& V_X \rightarrow V_X + {i \over 2} (\Lambda - {\bar \Lambda}) \ ,
\nonumber \\
&& M_k \rightarrow M_k + {1 \over 2} \epsilon_k \Lambda \ . \label{w5}   
\ea
Let us consider for simplicity the case of just one relevant twisted modulus
$M$. In this case, the mixed anomalies are cancelled provided the
following condition holds
\be
{\epsilon \over 4 \pi^2} = {A_1 \over s_1} = {A_2 \over s_2}= {A_3 \over
s_3} \ . \label{w6}
\ee
If the twisted field has a rather large vev $\langle m \rangle >> 
\langle s \rangle $, then the weak mixing angle at the scale of the 
symmetry breaking $M$ is determined by the relation
\be
\sin^2 \theta_W \simeq {s_2 \over s_1+s_2} = {A_2 \over A_1+A_2} = {1 \over
4} \ . \label{w7}
\ee
This value is rather close to the experimentally measured one at the
$M_Z$ scale $\sin^2 \theta_W \simeq 0.231$, as pointed long ago by
Weinberg and reemphasized recently in \cite{dk} in specific models.
The model we put forward here is the MSSM with an additional
abelian symmetry. It is therefore natural to analyse the RGE for the
gauge couplings and to explicitly find the energy scale at which
(\ref{w7}) holds.  A straightforward computation, with the one-loop
$\beta$-functions of the SM between $M_Z$ and an effective MSSM
threshold, $M_S$, and with the  MSSM $\beta$-functions above $M_S$
gives for the scale $\mu_0$ where(\ref{w7}) holds the expression
\be
\mu_0 \ = \ M_Z \left(\frac{M_Z}{M_S}\right)^{1/4} 
e^{{\pi \over 4} [{1 \over \alpha_Y (M_Z)}- {3 \over \alpha_2 (M_Z)}]} 
\ \simeq \ 90 - 150 \ TeV \ , \label{w10} 
\ee
a value within the energy range of the fundamental scale of our model .

We consider (\ref{w7}),(\ref{w10}) as an encouraging hint in order to
embed our model containing the additional lepton-like multiplets
into a fundamental theory. The relation $A_2=A_3$ in
(\ref{w3}) implies $s_2=s_3$ and therefore unification of the $SU(3)_c$
strong coupling with the $SU(2)_L$ gauge coupling at the string scale.
Even if this is an appealing feature, the logarithmic evolution
of gauge couplings \cite{gqw} cannot explain this strong-weak
unification.  There are several possibilities in order to overcome this
difficulty.  One possibility we mention here is the case
where the $SU(3)_c$ strong interaction brane (but not the electroweak
branes) experience a new large compact dimension in the TeV range
\cite{antoniadis}. The strong coupling will then start having a
power-law evolution \cite{tv,ddg} and it will unify with the weak
$SU(2)_L$ coupling at low energy. This is similar with the unification
mechanism proposed in \cite{ddg}, but uses the power-law evolution 
just for the strong gauge coupling.  Another possibility is that the
$SU(3)$ gauge field couples to additional moduli fields on the strong
coupling brane. The v.e.v. of the  additional (twisted or untwisted)
fields can then modify the strong coupling in a phenomenologically
successful, but unpredictible from our arguments, way.

\noindent
{\bf Acknowledgments.} We dedicate this work to our collaborator and
friend Stefan Pokorski, on the occasion of his 60th birthday. Stefan
greatly stimulated the interest for phenomenology of E.D. by
completing, through uncountable discussions over the last years, his
particle physics education. We would like to thank Christophe Grojean
and Stephane Lavignac for helpful discussions. Work supported in part
by the RTN European Program HPRN-CT-2000-00148.

%%%%%%%%%%%%%%%%%%%%%%%%%%%%%%%%%%%%%%%%%%%%%%%%%%%%%%%%%%%%%%%%%%%%%%

%%%%%%%%%%%%%%%%%%%%%%%%%%%%%%%%%%%%%%%%%%%%%%%%%%%%%%%%%%%%%%%%%%%%%%%%%%%
\end{document}